\documentclass[prd,twocolumn,superscriptaddress,showpacs,amssymb,amsmath,
amsfonts,aps,altaffilletter]{revtex4}

\usepackage{color}
\usepackage{times}
\usepackage{graphicx}
\usepackage{fancyhdr}
\usepackage{float}
\usepackage{ulem}
\usepackage[raggedright]{subfigure}
\usepackage{acronym, multirow}
\usepackage[style=altlist]{glossary}

\makeatletter
\makeatletter
\def\@fnsymbol#1{\ifcase#1\or * \or  $+$ \or  \$ \or \#  \or \dag \or \ddag \or
$\mathsection$ \or $ \mathparagraph$ \or $\|$  \or \textordfeminine \or \textbul
let   
\or ** \or $++$ \or  \$\$ \or \#\#  \or \dag\dag \or \ddag\ddag \or
$\mathsection\mathsection$ \or $ \mathparagraph\mathparagraph$ \or $\|\|$  \or 
\textordfeminine\textordfeminine \or \textbullet \textbullet \or *** \or $+++$ 
\or  \$\$\$ \or \#\#  \or \dag\dag \or \ddag\ddag \or
$\mathsection \mathsection\mathsection$ \or $ \mathparagraph 
\mathparagraph\mathparagraph$ \or $\|\|\|$  \or 
\textordfeminine\textordfeminine\textordfeminine \or 
\textbullet\textbullet\textbullet \or \else \@ctrerr\fi}
\makeatother

\def\thercsid{\relax}
\def\rcsid#1{\def\next##1#1{\def\thercsid{##1}}\next}
\rcsid$Id: paper.tex,v 1.179 2008/01/23 11:46:13 thomas Exp $
\newcommand\GW{gravitational wave~}

\renewcommand{\today}{\number\day\space\ifcase\month\or
  January\or February\or March\or April\or May\or June\or
  July\or August\or September\or October\or November\or December\fi
  \space\number\year}

\newacronym{BBH}{binary black hole}{}
\newacronym{BNS}{binary neutron star}{}
\newacronym{PBH}{primordial black hole}{}
\newacronym{SNR}{signal-to-noise ratio}{}
\newacronym{LIGO}{Laser Interferometer Gravitational-wave Observatory}{}
\newacronym{LSC}{LIGO Scientific Collaboration}{}
\newacronym{GRB}{gamma-ray bursts}{}

\begin{document}

\title{Search for gravitational waves from binary inspirals in S3 and S4 LIGO
data 
}

\affiliation{Albert-Einstein-Institut, Max-Planck-Institut f\"ur Gravitationsphysik, D-14476 Golm, Germany}
\affiliation{Albert-Einstein-Institut, Max-Planck-Institut f\"ur Gravitationsphysik, D-30167 Hannover, Germany}
\affiliation{Andrews University, Berrien Springs, MI 49104 USA}
\affiliation{Australian National University, Canberra, 0200, Australia}
\affiliation{California Institute of Technology, Pasadena, CA  91125, USA}
\affiliation{California State University Dominguez Hills, Carson, CA  90747, USA}
\affiliation{Caltech-CaRT, Pasadena, CA  91125, USA}
\affiliation{Cardiff University, Cardiff, CF24 3AA, United Kingdom}
\affiliation{Carleton College, Northfield, MN  55057, USA}
\affiliation{Charles Sturt University, Wagga Wagga, NSW 2678, Australia}
\affiliation{Columbia University, New York, NY  10027, USA}
\affiliation{Embry-Riddle Aeronautical University, Prescott, AZ   86301 USA}
\affiliation{Hobart and William Smith Colleges, Geneva, NY  14456, USA}
\affiliation{Inter-University Centre for Astronomy  and Astrophysics, Pune - 411007, India}
\affiliation{LIGO - California Institute of Technology, Pasadena, CA  91125, USA}
\affiliation{LIGO Hanford Observatory, Richland, WA  99352, USA}
\affiliation{LIGO Livingston Observatory, Livingston, LA  70754, USA}
\affiliation{LIGO - Massachusetts Institute of Technology, Cambridge, MA 02139, USA}
\affiliation{Louisiana State University, Baton Rouge, LA  70803, USA}
\affiliation{Louisiana Tech University, Ruston, LA  71272, USA}
\affiliation{Loyola University, New Orleans, LA 70118, USA}
\affiliation{Moscow State University, Moscow, 119992, Russia}
\affiliation{NASA/Goddard Space Flight Center, Greenbelt, MD  20771, USA}
\affiliation{National Astronomical Observatory of Japan, Tokyo  181-8588, Japan}
\affiliation{Northwestern University, Evanston, IL  60208, USA}
\affiliation{Rochester Institute of Technology, Rochester, NY 14623, USA}
\affiliation{Rutherford Appleton Laboratory, Chilton, Didcot, Oxon OX11 0QX United Kingdom}
\affiliation{San Jose State University, San Jose, CA 95192, USA}
\affiliation{Southeastern Louisiana University, Hammond, LA  70402, USA}
\affiliation{Southern University and A\&M College, Baton Rouge, LA  70813, USA}
\affiliation{Stanford University, Stanford, CA  94305, USA}
\affiliation{Syracuse University, Syracuse, NY  13244, USA}
\affiliation{The Pennsylvania State University, University Park, PA  16802, USA}
\affiliation{The University of Texas at Brownsville and Texas Southmost College, Brownsville, TX  78520, USA}
\affiliation{Trinity University, San Antonio, TX  78212, USA}
\affiliation{Universitat de les Illes Balears, E-07122 Palma de Mallorca, Spain}
\affiliation{Universit\"at Hannover, D-30167 Hannover, Germany}
\affiliation{University of Adelaide, Adelaide, SA 5005, Australia}
\affiliation{University of Birmingham, Birmingham, B15 2TT, United Kingdom}
\affiliation{University of Florida, Gainesville, FL  32611, USA}
\affiliation{University of Glasgow, Glasgow, G12 8QQ, United Kingdom}
\affiliation{University of Maryland, College Park, MD 20742 USA}
\affiliation{University of Michigan, Ann Arbor, MI  48109, USA}
\affiliation{University of Oregon, Eugene, OR  97403, USA}
\affiliation{University of Rochester, Rochester, NY  14627, USA}
\affiliation{University of Salerno, 84084 Fisciano (Salerno), Italy}
\affiliation{University of Sannio at Benevento, I-82100 Benevento, Italy}
\affiliation{University of Southampton, Southampton, SO17 1BJ, United Kingdom}
\affiliation{University of Strathclyde, Glasgow, G1 1XQ, United Kingdom}
\affiliation{University of Washington, Seattle, WA, 98195}
\affiliation{University of Western Australia, Crawley, WA 6009, Australia}
\affiliation{University of Wisconsin-Milwaukee, Milwaukee, WI  53201, USA}
\affiliation{Vassar College, Poughkeepsie, NY 12604}
\affiliation{Washington State University, Pullman, WA 99164, USA}
\author{B.~Abbott}\affiliation{LIGO - California Institute of Technology, Pasadena, CA  91125, USA}
\author{R.~Abbott}\affiliation{LIGO - California Institute of Technology, Pasadena, CA  91125, USA}
\author{R.~Adhikari}\affiliation{LIGO - California Institute of Technology, Pasadena, CA  91125, USA}
\author{J.~Agresti}\affiliation{LIGO - California Institute of Technology, Pasadena, CA  91125, USA}
\author{P.~Ajith}\affiliation{Albert-Einstein-Institut, Max-Planck-Institut f\"ur Gravitationsphysik, D-30167 Hannover, Germany}
\author{B.~Allen}\affiliation{Albert-Einstein-Institut, Max-Planck-Institut f\"ur Gravitationsphysik, D-30167 Hannover, Germany}\affiliation{University of Wisconsin-Milwaukee, Milwaukee, WI  53201, USA}
\author{R.~Amin}\affiliation{Louisiana State University, Baton Rouge, LA  70803, USA}
\author{S.~B.~Anderson}\affiliation{LIGO - California Institute of Technology, Pasadena, CA  91125, USA}
\author{W.~G.~Anderson}\affiliation{University of Wisconsin-Milwaukee, Milwaukee, WI  53201, USA}
\author{M.~Arain}\affiliation{University of Florida, Gainesville, FL  32611, USA}
% ...
\collaboration{The LIGO Scientific Collaboration, http://www.ligo.org}
\noaffiliation

\date[\relax]{ RCS \thercsid; compiled \today }
\pacs{95.85.Sz, 04.80.Nn, 07.05.Kf, 97.80.-d}

\begin{abstract}\quad
We report on a search for gravitational waves from the coalescence of 
compact binaries during the third and fourth LIGO science runs. The
search focused on gravitational waves generated during the
inspiral phase of the binary evolution. In our analysis, we considered three
categories of compact binary systems, ordered by mass: (i) primordial black
hole binaries with masses in the range $0.35 M_\odot < m_1,m_2 < 1.0 M_\odot$,
(ii) binary neutron stars with masses in the range $1.0 M_\odot < m_1,m_2 < 3.0
M_\odot$, and (iii) binary black holes with masses in the range $3.0
M_\odot< m_1,m_2 < m_{\rm max}$ with the additional constraint
$m_1+m_2 < m_{\rm max}$, where $m_{\rm max}$ was set to $40.0 M_{\odot}$ and
$80.0 M_\odot$ in the third and fourth science runs, respectively.
Although the detectors could
probe to distances as far as tens of Mpc, no gravitational-wave
signals were identified in the 1364 hours of data we analyzed. Assuming a binary
population with a Gaussian distribution around $0.75\textrm{--}0.75 M_{\odot}$,
$1.4\textrm{--}1.4 M_{\odot}$, and $5.0\textrm{--}5.0 M_{\odot}$, we derived
$90\%$-confidence upper limit rates of $4.9~\textrm{yr}^{-1} L_{10}^{-1}$ for
primordial black hole binaries, $1.2~\textrm{yr}^{-1} L_{10}^{-1}$ for binary
neutron stars, and $0.5~\textrm{yr}^{-1} L_{10}^{-1}$ for stellar mass binary
black holes, where $L_{10}$ is $10^{10}$ times the blue light luminosity of the
Sun. 
\end{abstract}


\maketitle

\section{Overview}\label{sec:overview}

While gravitational radiation has not yet been directly detected, 
observations of the orbital decay of the first binary pulsar PSR
B1913+16 \cite{Taylor:1982,Taylor:1989} have provided
significant indirect evidence for their existence since the late
eighties.
Indeed, observations have revealed a gradual inspiral to within about
0.2
percent of the rate expected from the emission of gravitational
radiation
\cite{weisberg:2004}.
As orbital energy and angular momentum are carried away by
gravitational radiation,  the two compact objects in a binary system
become more
tightly bound and orbit faster until they eventually merge. The
gravitational
wave signals emitted by the merging of binary systems made of primordial black holes,
neutron
stars, and/or stellar mass black holes can be detected by ground-based
detectors. The detection rate depends on the merger rate, which in turn
depends on the rate of ongoing star
formation within
LIGO's detection volume, described in greater detail in
\cite{LIGOS3S4Galaxies} and as measured by the net blue luminosity
encompassed in that volume (see Sec.~\ref{sec:ul}).

Several direct and indirect methods can be applied to infer the merger
rate expected per unit $L_{10}$,   where $L_{10}$ is $10^{10}$ times
the
blue solar luminosity.
%
Merger rates for {\BNS}
systems can be directly inferred from the four systems observed as
binary pulsars that
will merge in less than a Hubble time; the basic methodology was
originally applied by 
\cite{Phinney:1991ei,Narayan:1991}.  The current estimates based 
on all known {\BNS} suggest that  the merger rate lies
in the range $10\textrm{--}170 \times 10^{-6} \textrm{yr}^{-1}
L_{10}^{-1}$
\cite{Kalogera:2004nt,Kalogera:2004tn}. This range is at 95\% 
confidence for a specific model of the Galactic population (model \#6
in the references), which represents our current understanding of the
radio pulsar luminosity function and their Galactic spatial
distribution.  The most likely rate for the same model is 
$50 \times 10^{-6} \textrm{yr}^{-1} L_{10}^{-1}$
\cite{Kalogera:2004nt,Kalogera:2004tn}.
The estimated BNS merger rate makes the detection of a signal from
such an
event unlikely, though possible, with the current generation of
gravitational-wave detectors.  
%
In contrast,
there is no direct astrophysical evidence for the existence of {\BBH}
or black hole/neutron star binaries, but they are predicted to exist
on the
basis of our current understanding of compact object formation and
evolution.
The search for gravitational waves emitted by {\BBH} systems is
particularly interesting since it would provide direct observation of
these
systems.  
Merger rate estimates are currently obtained from
theoretical population
studies of binaries in galactic fields 
\cite{ADM:Tut93,1995ApJ...440..270B,1998AA...332..173P,1999ApJ...526..152F,ADM:Vos2003,Belczynski:2002,StarTrack2,lrr-2006-6}
or in dense stellar clusters
\cite{PortegiesZwart:2000,2006ApJForS3S4Joint,imbhlisa-2006}.   
Because these studies differ significantly in their assumptions and
methodology, it is difficult to assess \emph{all}  the literature and
assign
relative likelihoods to merger different merger rates for black hole
binaries.
However, in the case of field binaries, estimates for the relative
likelihood  can be obtained by  widely exploring several of the
parameters
of the population models, while ensuring those models reproduce the
{\BNS} merger rates derived from the observed sample
\cite{OShaughnessy:2005,Oshaughnessy:2006b}. Based on this study, the
merger rates for {\BBH} and black hole/neutron star binaries
are found to lie in the ranges (at 95\% confidence) 
 $0.1-15\times 10^{-6}\textrm{ yr}^{-1} L_{10}^{-1}$ and 
       $0.15-10\times 10^{-6}\textrm{ yr}^{-1} L_{10}^{-1}$
       respectively, with most likely merger rates of 
$0.6\times 10^{-6}\textrm{ yr}^{-1} L_{10}^{-1}$ and 
$1.3\times 10^{-6}\textrm{ yr}^{-1} L_{10}^{-1}$.
Although drawn from a single study, the simulations cover such a
uniquely wide parameter space that these rate ranges are
consistent with the existing literature on {\BBH} and black
hole/neutron star merger rates.
It has also been discussed in the literature that some fraction of all
dense clusters may form many inspiraling 
{\BBH}; although the current rate predictions are considered highly
uncertain and the systematic uncertainties are not yet understood,
rates as high as a few events per year detectable by initial LIGO have
been reported
\cite{PortegiesZwart:2000,2006ApJForS3S4Joint,imbhlisa-2006}.   
%
Furthermore, indirect evidence suggests that
short, hard {\GRB}s could be associated with the coalescence of a
{\BNS} or a 
black hole/neutron star binary.  Recent estimates suggest that the
rates of
these events could be in excess of about $1 \times 10^{-6} \textrm{
  yr}^{-1}
L_{10}^{-1}$ \cite{Nakar:2006}. 
There may also exist sub-solar-mass black hole
binary systems, with component objects that could have formed in the
early
universe and which contribute to galactic dark matter halos
\cite{Alcock:2000ph};
we refer to such lower-mass compact binary coalescences as {\PBH}
binaries.

The {\LIGO} Scientific Collaboration (LSC) operates four
interferometric detectors. Three of these are from the U.S.
{\LIGO} project \cite{LIGOS1instpaper, Barish:1999}, two of them, with 4~km and
2~km long arms, are co-located in Hanford, WA (called H1 and H2, respectively)
and a third detector, with 4~km long arms, is located in Livingston, LA (called
L1). The LSC also operates the British-German GEO~600 detector
\cite{Luck:1997hv}, with 600~m long arms that is located near
Hannover, Germany. Only data from the LIGO detectors were used
in this analysis, however, due to the relative sensitivity of the detectors.

\begin{figure}[tbp]
\includegraphics[width=3.4in,angle=0, height=2.5in]{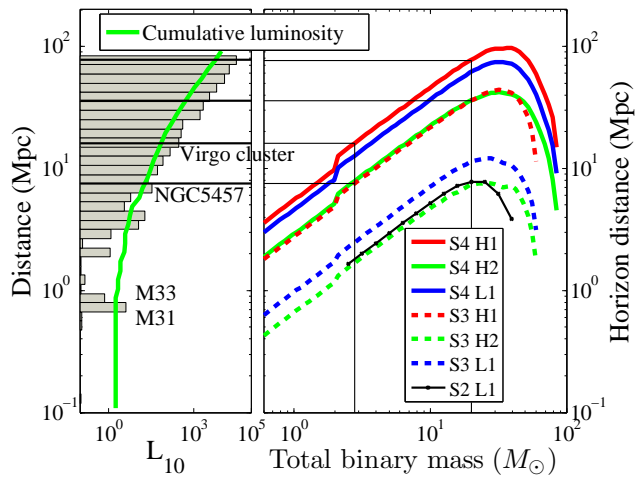}
\caption{Blue-light luminosities and horizon distances for LIGO's 
observatories. In the left panel, the horizontal bars represent the
non-cumulative \textit{intrinsic} blue-light luminosity of the  galaxies or clusters
within each bin of \textit{physical} distance, as obtained from a standard astronomy
catalog~\cite{LIGOS3S4Galaxies}. Some bins are identified by the dominant
contributor galaxy or cluster. The merger rate of binaries within a galaxy or cluster
is assumed to scale with its blue-light luminosity. 
The detectability of a binary
depends on the \textit{effective} distance between the source and detector, which
is dependent on both the physical distance separating them and their relative
orientation (see Eq.~\ref{eq:Deff}). The solid line shows the cumulative
blue-light luminosity as a function of effective distance (hereafter, the
\textit{effective} cumulative blue-light luminosity), of the binary sources which would
be observed by the LIGO detectors if they had perfect detection efficiency (i.e.,
all binaries are detectable). Explicitly, a binary in M31 (at a physical
distance of $0.7 \textrm{ Mpc}$) with an effective distance of $5 \textrm{ Mpc}$ will contribute to the
\textit{intrinsic} luminosity at $0.7 \textrm{ Mpc}$ and will contribute to the
\textit{effective} cumulative  luminosity at a distance of $5 \textrm{ Mpc}$. Although a binary
will have slightly different orientation with respect to each LIGO observatory and
therefore slightly different effective distances, the difference in the \textit{effective}
cumulative luminosity shown on this plot would not be distinguishable. The
\textit{effective} cumulative luminosity starts at $1.7~L_{10}$ (Milky Way contribution),
and begins increasing at a distance of $\sim 1\textrm{ Mpc}$, with the contribution of nearby
galaxies M31 and M33. The cumulative luminosity observable by our search (not
shown), as expressed in Eq.~\ref{eq:cl}, depends also on the detection efficiency of our
search (see Fig.~\ref{fig:EffvsDist}) and will be less than the effective cumulative
luminosity. In the right panel, the curves represent the horizon distance in each LIGO
detector as a function of total mass of the binary system, during S3 (dashed lines) and S4
(solid lines). We also plot the horizon distance of L1 during S2. The sharp drop of
horizon distance around a total mass of 2$M_\odot$ is related to a different lower
cut-off frequency, $f_L$, used in the PBH binary search and the BNS/BBH searches. The
$f_L$ values are summarized in Table~\ref{tab:space}. The high cut-off frequency occurs
at the last stable orbit. The horizon distance for non equal mass systems scales by
a factor $\sqrt{4 m_1 m_2}/(m_1+m_2)$.}
\label{fig:LumvsDist}
\end{figure}

We report on a search for gravitational waves emitted by
coalescing compact binaries in the data taken by the {\LIGO} detectors in
late 2003 (Oct 31, 2003-Jan 9, 2004) and early 2005 (Feb 22, 2005-March
24, 2005) which correspond to the third (S3) and fourth (S4) science runs,
respectively. During S3 and S4, the LIGO detectors were
significantly more sensitive than in our previous science runs \cite{LIGOS1iul,
LIGOS2iul,LIGOS2macho,LIGOS2bbh}. This improvement can be quantified in terms
of the inspiral \textit{horizon distance} of each detector which is
defined as the distance at which an optimally located and oriented binary
system would give expected {\SNR} equal to 8.  For instance, H1, the most
sensitive detector during S4, had 
horizon distance averaged over the duration of the run of $5.7 \textrm{ Mpc}$,
$16.1 \textrm{ Mpc}$, and
$77.0 \textrm{ Mpc}$, 
for a $0.5\textrm{--}0.5 M_{\odot}$, $1.4\textrm{--}1.4 M_{\odot}$, and
 $10\textrm{--}10 M_{\odot}$ systems,
respectively. Consequently, during S3 and S4, the detectors were
sensitive enough to detect inspiral signals from hundreds of galaxies as shown
in Fig.~\ref{fig:LumvsDist}.

The paper organization is as follows. In Sec.~\ref{sec:pipeline}, we
briefly describe the data analysis pipeline and present the parameters
 used in the S3 and S4 science runs.  In particular, Sec.~\ref{sec:search}
describes the division of the search into 3 categories of  binaries: {\PBH}
binary, {\BNS}, and {\BBH} inspirals. In Sec.~\ref{sec:results}, we present the
results of the search, including the accidental rate estimates
and loudest
candidates found from the different science runs and categories of binary
systems that we considered. Finally, Sec.~\ref{sec:ul} describes the upper
limits set by this analysis.

\section{The data analysis pipeline}\label{sec:pipeline}

The analysis pipeline used to search  the S3 and S4 data received
substantial improvements over the one used in our previous 
searches~\cite{LIGOS1iul,LIGOS2iul,LIGOS2macho,LIGOS2bbh}.
The pipeline is fully described in a set of companion
papers~\cite{findchirppaper,LIGOS3S4Tuning,BBCCS:2006}; this section introduces the
aspects of our analysis methodology that are essential to comprehend the search
and the final upper limit results. We emphasize the
differences between the BBH search and the PBH binary/BNS searches.

\subsection{Coincident data and  time analyzed}\label{sec:times}
The first step of the analysis pipeline is to prepare a list of
time intervals represented by a start and end time, during which
at least two detectors are operating nominally. Requiring coincident signals
from two or more detectors reduces the accidental rate by several orders of
magnitude and increases our detection confidence.

In S3, we required both Hanford detectors to be operating; analyzed times
belonged either to triple H1-H2-L1 or double H1-H2 coincident times. In S4,
times when H1 was operating but
H2 was not (and vice-versa) were also analyzed, therefore all permutations
of double coincident times were possible, in addition to the triple coincident
times. 
The breakdown of times
analyzed, common to all searches, is given in Table~\ref{tab:analysedtimes}. A
fraction of these times (about 9\%), \textit{playground} times,  was used to tune
the search parameters. This tuning was
performed in order to suppress  background triggers originating from
instrumental noise so as to efficiently detect the gravitational wave signals
(measured using simulated injections, as described in
Section~\ref{sec:background}). In order to avoid potential bias, upper limits
(Section~\ref{sec:ul}) are derived using the non-playground data only. However,
candidate detections are drawn from the full data set.

We compiled a list of time intervals when the detectors had poor data
quality~\cite{LIGOS3S4Tuning, Vetoes}.
In S3, this selection discarded  5\% of H1/H2 as a result of high
seismic noise and 1\% of L1 data as a result of data acquisition overflow. In
S4, 10\% of H1/H2 data was discarded mostly
due to transients  produced when one Hanford detector was operating but the 
other was not. 
A gravitational wave arriving during one of the vetoed times
could, under certain conditions, still be detected and validated. However,
neither playground times nor vetoed times are included when computing the upper
limits presented in Section~\ref{sec:ul}.

\begin{table}
\caption{Times analyzed when at least two detectors were operating. The times
in parentheses exclude \textit{playground} times, which represents
about 9\% of the data and is used to tune
the search.}
\begin{center}
\begin{ruledtabular}
\begin{tabular}{lcc}
 & S3 & S4 \\\hline
H1-H2-L1 times & 184 (167) hrs & 365 (331) hrs\\
H1-H2 times & 604 (548) hrs & 126 (114) hrs\\
H1-L1 times & -- & 46 (41) hrs\\
H2-L1 times & -- & 39 (35) hrs\\\hline
Total times & 788 (715) hrs & 576 (521) hrs
\end{tabular}
\end{ruledtabular}
\end{center}
\label{tab:analysedtimes}
\end{table}

\subsection{Filtering}\label{sec:effDist}

In the adiabatic regime of binary inspiral, gravitational wave
radiation is modeled accurately. We make use of a variety of approximation
techniques
\cite{Blanchet:1995ez,Blanchet:1995,Blanchet:1996pi,Blanchet:2001ax,
Blanchet:2004ek,Damour:1998zb,BuonannoDamour:1999,BuonannoDamour:2000,
Damour:2000zb} which rely, to some extent, on the slow motion of the compact
objects which make up the binary. We can represent the known waveform by
\begin{equation}\label{eq:h}
h(t) = \frac{1 {\rm Mpc}}{D_{\rm eff}} A(t)\cos{\left(\phi(t) - \phi_0 \right)}
\end{equation}
where $\phi_0$ is some unknown phase,  and the functions $A(t)$ and
$\phi(t)$ depend on the masses and spins of the binary.
Although spin effects can be taken into account
\cite{S3_BCVSpin}, they are estimated to be negligible over much of the
mass range explored in this search and will be neglected here. 
Since the \GW signal we are searching for is known, the matched filtering
method of detection constitutes the cornerstone of our analysis. In both {\PBH}
binary and
{\BNS} searches, we use physical {template families} based on second
order restricted post-Newtonian waveforms in the stationary-phase approximation
\cite{Blanchet:1996pi, Droz:1999qx}. In the {\BBH} search, we use a
phenomenological template family \cite{BuonannoChenVallisneri:2003a} so as to
palliate uncertainties in the gravitational-wave templates, which become
significant in the LIGO band for higher mass systems. The template matched
filtering will identify the masses and coalescence time of the binary but not
its physical distance $D$. The signal amplitude received by the
detector depends on the detector response functions $F_+$ and
$F_\times$, and the inclination angle of the source $\iota$, which are unknown.
We can only obtain the \textit{effective distance} $D_{\rm eff}$, which
appears in Eq.~(\ref{eq:h})  defined as~\cite{thorne.k:1987}:
\begin{equation}\label{eq:Deff}
D_{\rm eff} = \frac{D}{\sqrt{F_+^2 (1 + \cos^2\iota)^2/4 + F_\times^2
( \cos\iota)^2}} \; .
\end{equation}
The effective distance of a binary may be larger than its physical
distance.

\subsection{Inspiral search parameters}\label{sec:search}
We searched for {\PBH} binaries with component masses between $0.35~M_\odot$ and
$1~M_\odot$, and {\BNS} with component masses between $1~M_\odot$ and
$3~M_\odot$. We also searched for {\BBH} systems with component masses between
$3~M_\odot$ and $m_{\rm max}$, 	where $m_{\rm max}$ was set to
$40~M_\odot$ and $80~M_\odot$ in S3 and S4, respectively. In addition, the total
mass of the systems was also constrained to be less than $m_{\rm max}$.
The larger mass range in S4 is due to improvement of the detector sensitivities
at low frequency. This
classification of binaries into three categories was driven primarily by
technical
issues in the data analysis methods. In particular, the waveforms differ
significantly from one end of the mass scale to the other: gravitational
waves from lower mass binaries last tens of seconds in the LIGO band and require
more templates to search for them, as compared to the higher mass
binaries (see Table \ref{tab:space}).

For each search, we filtered the data through 
template banks designed to cover the corresponding range of component masses.
The template banks are generated for each detector and each 2048-second data
stretch so as to take into account fluctuations of the power spectral
densities. In the {\PBH} binary and {\BNS} searches, the algorithm devoted to
the template bank placement \cite{BBCCS:2006} is identical to the one used in
previous searches \cite{LIGOS2iul,LIGOS2macho}. In the
{\BBH} search, we used a phenomenological bank placement similar to 
the one used in the S2 BBH search~\cite{LIGOS2bbh}. The spacing between
templates gives at most 5\% loss of SNR in
the {\PBH} binary and {\BBH} banks, and 3\% in the {\BNS} bank.  The
average number of templates needed to cover the parameter space of each binary
search are shown in Table~\ref{tab:space}, and are indicative of the
relative computational cost of each search.

\begin{table}[htdp]
\caption{The target sources of the search. The second and third columns
show the mass ranges of the binary systems considered. The fourth column
provides the lower
cut-off frequency, $f_L$, which set the length of the templates, and the
fifth column gives the average number of templates needed, $N_{\rm b}$.
The last column gives the longest waveform duration, $T_{\rm
max}$.}
\begin{ruledtabular}
      \begin{tabular}{crrrrrr}
	
  	\multicolumn{1}{c}{}
	& $m_{\rm min}(M_\odot)$
	& $m_{\rm max}(M_\odot)$	
	& $f_L$(Hz)
	& $N_{\rm b}$ & $T_{\rm max}$(s)\\\hline
	S3,S4 PBH & 0.35 & 1.0 & 100 & 4500&22.1\\\hline	
	S3 BNS & 1.0 & 3.0 &70 & 2000&10.0\\
	S4 BNS & 1.0 & 3.0 &40 & 3500&44.4\\\hline
	S3 BBH & 3.0 & 40.0 &70 & 600& 1.6\\
	S4 BBH & 3.0 & 80.0  &50 & 1200& 3.9\\
      \end{tabular}
\end{ruledtabular}
\label{tab:space}
\end{table}

For each detector, we construct a template bank which
we use to filter the data from the gravitational wave channel.
Each template produces an {\SNR} time series, $\rho(t)$. We only keep
stretches of $\rho(t)$ that exceed a preset  threshold (6.5 in the PBH
binary and BNS searches and $6$ in the {\BBH} case). Data reduction
is necessary to cope with the large rate of triggers that are mostly due
to noise transients. First, \emph{each} {\SNR} time series is clustered
using
a sliding window of 16~s as explained in \cite{findchirppaper}. Then, 
 surviving triggers from all templates in the bank are clustered, so that only
the loudest template trigger is kept in fixed intervals of 10~ms
({\PBH} binary and
{\BNS}) or 20~ms ({\BBH}). These triggers constitute the output of the first
inspiral filtering step. To further suppress false triggers, we require
 additional checks such as coincidence in time in at least two detectors, as
described below.

\subsection{Coincidence parameters and combined SNR}\label{sec:combined}

In the {\PBH} binary and {\BNS} searches, we require coincidence in time, chirp
mass $\mathcal{M}_c=((m_1m_2)^3/(m_1+m_2))^{1/5}$, and symmetric mass ratio
$\eta={m_1 m_2}/(m_1+m_2)^2$. In the {\BBH} search, we
require coincidence in time, and the two phenomenological parameters $\psi_0$
and $\psi_3$, which correspond to first approximation to
$\mathcal{M}_c$ and $\eta$
 parameters, respectively (see \cite{BuonannoChenVallisneri:2003a,LIGOS2bbh}).
After the first inspiral filtering
step, which does not use any computationally expensive vetoing
methods such as a $\chi^2$ veto~\cite{Allen:2004}, we apply
coincidence windows with parameters that are summarized in
Table~\ref{tab:coinc}. Then, in the {\PBH} binary and {\BNS} searches, we employ
an hierarchical pipeline, in which coincident triggers are
re-filtered, and the $\chi^2$ veto is
calculated. Finally, trigger selection and coincidence requirements are
re-applied. In the {\BBH} search, no $\chi^2$ test is used because the
waveforms have very few cycles in the LIGO detector frequency band. The
coincident triggers from the first filtering step constitute the output of the
BBH search. The coincident triggers from the second filtering step constitute
the output of the PBH binary and BNS searches.

\begin{table}[htdp]
\caption{\label{tab:coinc} Summary of the S3 and S4 coincidence
windows. The second column gives the time coincidence windows; we also need to
account for the maximum light travel time between detectors (10~ms between
the L1
and H1/H2 detectors). The third column gives the chirp mass (PBH and BNS
searches), and $\psi_0$ coincidence windows (BBH search). In the
S4 BBH case, $\Delta\psi_0$  corresponds to about 1/15 of the $\psi_0$ range
 used in the template bank. The $\eta$ (PBH and BNS searches) and $\psi_3$ (BBH
search) parameters (last column) are not measured precisely enough to be
used in coincidence checks, except in
the
S4 BBH search.}
\begin{ruledtabular}
\begin{tabular}{lccc}
&  $\Delta T~{\rm (ms)}$ & $\Delta\mathcal{M}_c~(M_{\odot})$    & $\Delta \eta$
\\
S3/S4 PBH		& $4\times2$   & $0.002\times2$ & - \\
S3/S4 BNS		& $5\times2$   & $0.01\times2$  & - \\\hline

& $\Delta T~{\rm (ms)}$ & $\Delta\psi_0$    &
$\Delta\psi_3$ \\
S3 BBH		& $25\times2$ 	        & $40000\times2$    & - \\
S4 BBH		& $15\times2$ 	        & $18000\times2$    &$800\times2$
\\
\end{tabular}
\end{ruledtabular}
\end{table}

In the {\PBH} binary and {\BNS} searches, the $\chi^2$~test
provides a measure of the quality-of-fit of the
signal to the template. We can define an effective {\SNR}, $\rho_{{\rm eff}}$,
that combines $\rho$ and the $\chi^2$ value, calculated for the
same filter, by
\begin{equation}\label{e:rhoeff}
\rho_{\rm eff}^2 =
\frac{\rho^2}{\sqrt{\left(\frac{\chi^2}{2p-2 }\right)\left(1+
\frac{\rho^2}{250}\right)}}, 
\end{equation}
where $p$ is the number of bins used in the $\chi^2$ test; the specific
value of $p=16$ and the parameter $250$ in Eq.~(\ref{e:rhoeff}) are chosen
empirically, as justified in \cite{LIGOS3S4Tuning}. We expect $\rho_{\rm eff}\sim \rho$
for true signals with relatively low SNR, and low effective SNR for noise
transients. Finally, we assign to each coincident trigger a combined SNR,
$\rho_c$, defined by 
\begin{equation}
\label{eq:rhocbns}
(\rho_c)^2_{\rm BNS, PBH}=\sum_i^N \rho_{\rm eff, i}^2 \;, 
\end{equation} 
where $\rho_{\rm eff,i}$ is the effective SNR of the trigger ${\rm i^{th}}$
detector (H1, H2 or L1).

In the {\BBH} search, no $\chi^2$ test is calculated. Therefore  effective
{\SNR} cannot be used. Furthermore, the combined SNR defined in
Eq.~(\ref{eq:rhocbns}) does not represent a constant background trigger
statistic. Instead, we combine the SNRs from coincident triggers using a
\textit{bitten-L} statistic similar to the method used in S2 BBH search
\cite{LIGOS2bbh}, as justified in \cite{LIGOS3S4Tuning}.

Finally, for each type of search, the coincident triggers are 
clustered within a 10~s window ({\BNS} and {\BBH} searches) or 22~s window
({\PBH} binary search), distinct from the clustering mentioned in 
Sec.~\ref{sec:search}. The final coincident triggers constitute
the output of the pipeline---the \emph{in-time} coincident triggers.  

\section{Background and Loudest Candidates }\label{sec:results}
\subsection{Background}\label{sec:background}
To identify gravitational-wave event candidates, we need to estimate 
the probability of in-time coincident triggers arising from accidental
coincidence of noise triggers, which constitute our background, by
comparing the combined {\SNR} of in-time coincident triggers with the
expected background (with same or higher combined {\SNR}). In each search, we estimate 
the background  by repeating the analysis with the triggers from
each detector shifted in time relative to each other. In the three searches, we
used 50 time-shifts forward and the same number backward for the background
estimation, taking these as 100 experimental trials with no true signals to be expected in the coincident data set.
Triggers from H1 were not time-shifted, triggers from H2 were shifted 
by increments of $10 \textrm{ s}$, and triggers from L1 by $5 \textrm{ s}$. 

\begin{figure}[htdp]
{\label{fig:rhoeffscatterplot}
\includegraphics[width=3in,height=2.5in]{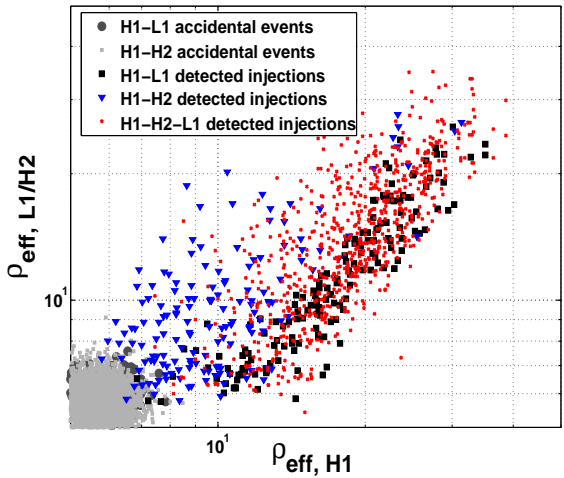}}
\caption{Accidental events and detected simulated injections. This plot shows
the distribution of effective SNR, $\rho_{\rm eff}$, as defined in
Eq.~(\ref{e:rhoeff}), for time-shifted coincident triggers and
detected  simulated injections (typical S4 BNS result). Some of the injections
are detected in all three detectors but no background triggers are found in
triple-coincidence in any of the 100 time shifts performed. The  H1-L1 and
H1-H2 time-shifted coincidence triggers have low effective SNR (left-bottom
corner).}
\label{fig:scatterplot}
\end{figure}

The time-shifted triggers are also used to explore the differences between noise and signal
events in our multi-dimensional parameter space. This comparison is performed
by adding simulated signals to the real data,  analyzing them with the same
pipeline, and determining the efficiency for detection of injected signals above
threshold. This procedure allows us to tune all aspects of
the pipeline on representative data without biasing our upper
limits. The general philosophy behind this tuning process is not to
perform aggressive cuts on the data, but rather to perform loose cuts and assess
our
confidence in a candidate by comparing where it lies in the multi-dimensional
parameter space of the search with respect to our expectations from background. 
The details of this tuning process are described in detail in a companion paper
\cite{LIGOS3S4Tuning}.  A representative scatter plot of the time-shifted
triggers and
detected simulated injections is shown in Fig.~\ref{fig:scatterplot} (S4 BNS
case). This plot also shows how the effective SNR statistic, which was used in
the PBH binary and BNS searches, separates background triggers from
simulated signals (with SNR as low as 8).

\begin{figure*}[htdp]
\includegraphics[width=0.33\textwidth]{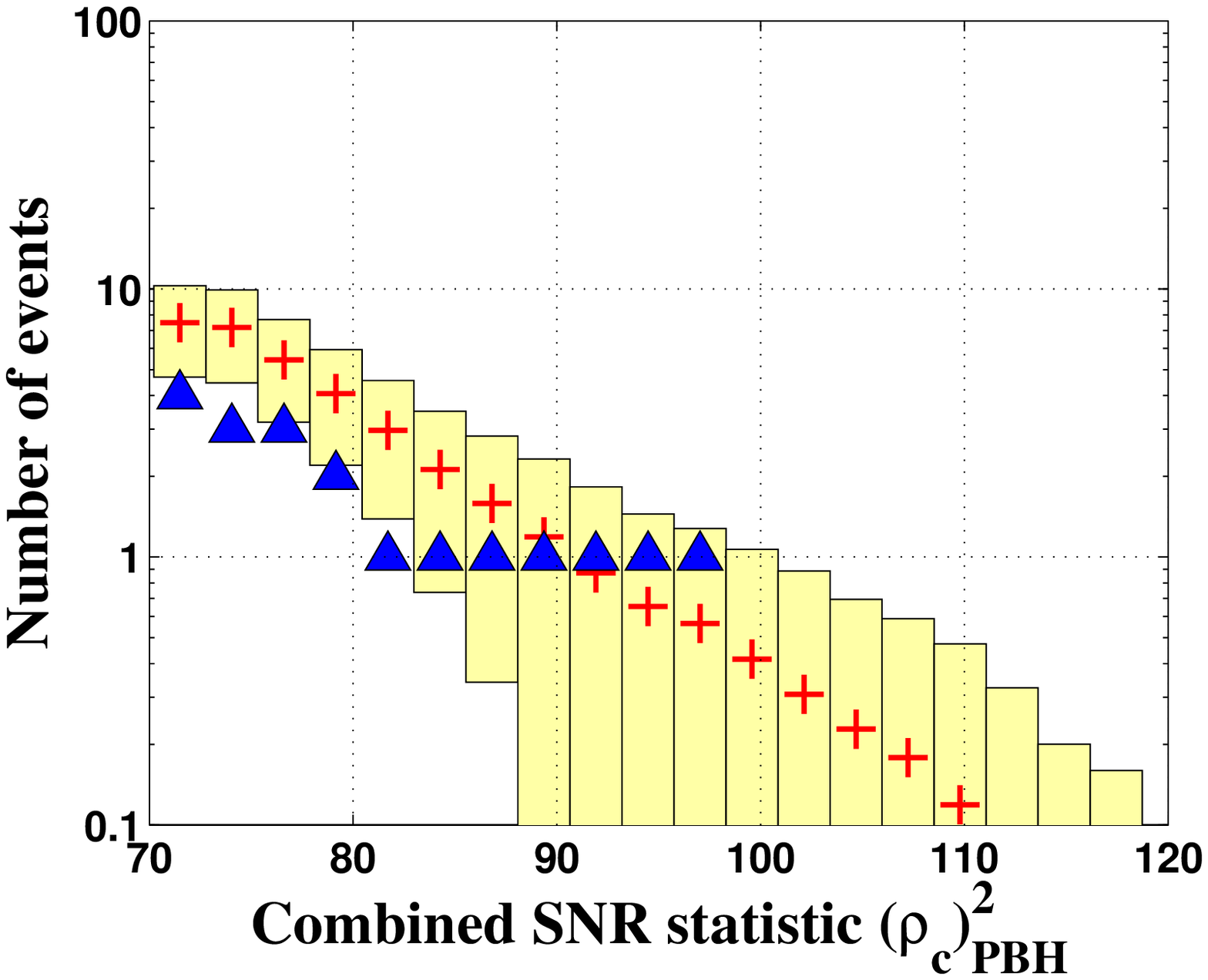}
\includegraphics[width=0.33\textwidth]{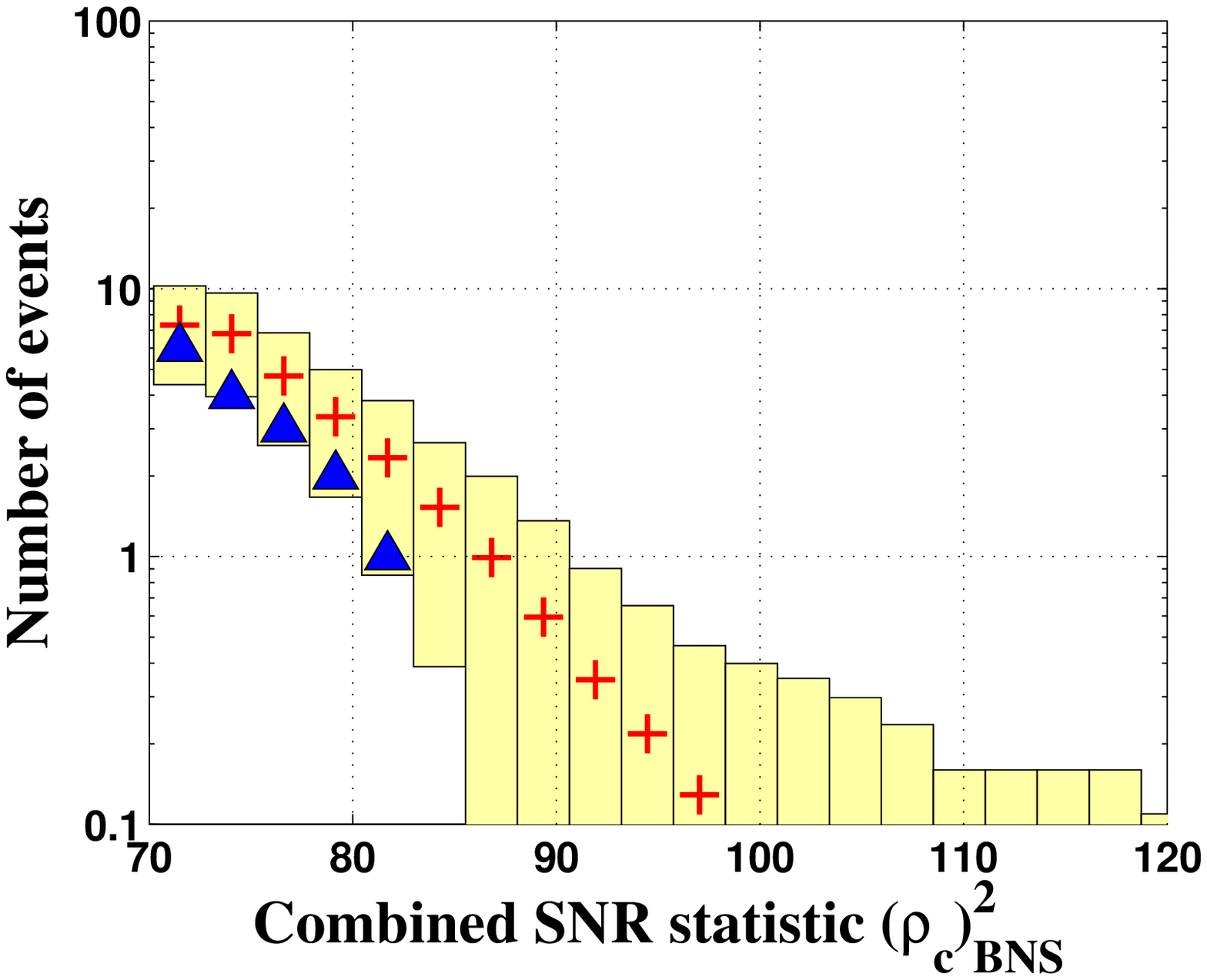}
\includegraphics[width=0.33\textwidth]{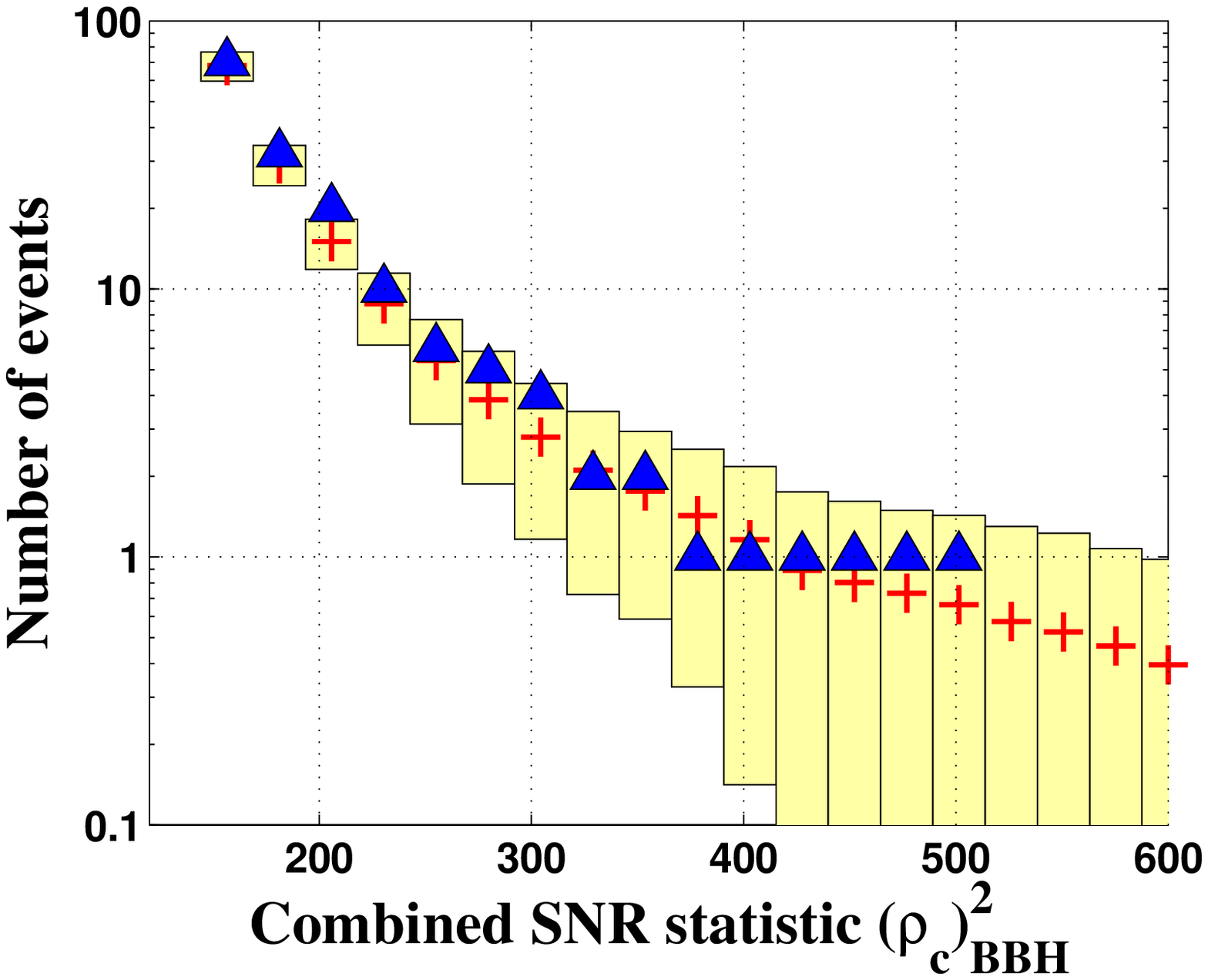}
\caption{Cumulative histograms of the combined SNR, $\rho_c$, for in-time
coincident candidates events (triangles) and estimated background from
accidental coincidences (crosses and 1 standard-deviation ranges), for the
S4 PBH binary (left), S4 BNS (middle) and S4 BBH (right) searches. In each
search, the loudest candidate (found in non-playground time) corresponds to an
accidental coincidence rate of about 1 during the entire S4 run.}
\label{fig:fgvsbg}
\end{figure*}

\subsection{Loudest candidates}\label{sec:loudest}
All searches had coincident triggers surviving at the end
of the pipeline. In order to identify a
gravitational wave event, we first compare the number of in-time coincident
triggers with the background estimate as a function of $\rho_c$. In S4, in-time
coincident triggers are
consistent with the background estimate in the three searches (see
Fig.~\ref{fig:fgvsbg}). Similar results were obtained in S3 PBH binary and S3
BNS searches. However, in the S3 BBH search (not shown), one event clearly lies
above expectation (in section~\ref{sec:bbh}, we explain why this candidate is
not 
a plausible gravitational wave detection). The criterion we used to 
identify detection candidates which exceed expectation is to associate them
with 
a probability $P_{\rm B}(\rho)$ that all background
events have a combined SNR smaller than $\rho$.  $P_{\rm B}(\rho)$
is calculated as the fraction of the 100 time-shifted experiments in
which all triggers have smaller combined SNR than $\rho$. 
A candidate with a large $P_{\rm B}$ is considered a plausible
gravitational wave event.
If this is the case and/or a candidate lies above expectation we
carefully scrutinize the data in the gravitational-wave channel and 
in auxiliary channels for possible instrumental noise that could 
produce an unusually loud false trigger. We also investigate the astrophysical
likelihood of the templates that best match the candidate in the different
detectors (e.g., the ratio of effective distances obtained in different
observatories). In addition, irrespective of the outcome of the comparison
between in-time and time-shifted coincidences, in-time coincident triggers with
the highest $\rho_c$ values are  also followed up.

The loudest coincident triggers found in each of the searches are listed in
Table~\ref{tab:Candidates}. Below, we briefly describe the reason(s) why we
rejected the loudest candidates found in the three searches performed on the S4
run. These loudest events are used for the upper limit calculation
(Sec.~\ref{sec:ul}). We also describe the loudest event found in the
S3 BBH search mentioned above.

\subsubsection{Primordial black hole binaries}

There were no {\PBH} binary candidates found in coincidence in all three
detectors with {\SNR} above the threshold of $6.5$; nor were there accidental
triple coincidences found in any of the 100 time-shifted runs. This means that
had there been a triple-coincident candidate, there would be less than a 1\%
probability of it being a background event ($P_{\rm B} \gtrsim 0.99$). 
A cumulative histogram of the combined SNR of the loudest in-time coincident
triggers in the S4 PBH search is shown in the leftmost plot of
Fig.~\ref{fig:fgvsbg}. The loudest S4 coincident trigger, with $\rho_{\rm
eff}= 9.8$, was found in coincidence in H1 and L1. We observed equally loud or
louder events in 58\% of the 100 time-shifted coincidence experiments. 
We found that this trigger was produced by a strong seismic transient
at Livingston, causing a much higher SNR in the L1 trigger than in the H1
trigger; we found many background triggers and some missed simulated
injections around the time of this event. As shown in
Table~\ref{tab:Candidates}, the
candidate also has significantly different effective distances in H1 ($7.4
\textrm{ Mpc}$) and L1
($0.07 \textrm{ kpc}$), because of the much larger SNR in L1: although not
impossible, such high ratios of effective distances are highly unlikely. Tighter
signal-based vetoes under development will eliminate these triggers in future
runs.

\subsubsection{Binary neutron stars}

Just as in the {\PBH} binary search, no triple coincident candidates or
time-shifted triple coincident candidates were found in the
{\BNS} search. In-time coincident triggers were found in pairs of
detectors only. We show in Fig.~\ref{fig:fgvsbg} (middle) the comparison of  the
number of coincident triggers  larger than a given $\rho_c$ with the
expected background  for S4.  The loudest coincident trigger was an
H1-L1 coincidence, consistent with estimated background, with $\rho_c=9.1$ and a
high probability of being a background trigger (See Table \ref{tab:Candidates}).

\subsubsection{Binary black holes} \label{sec:bbh}
Due to the absence of a $\chi^2$ waveform consistency test, the
BBH search suffered higher background trigger rates than the PBH binary and BNS
searches, and yielded candidate events found in triple coincidence, both in S3
and S4. All triple coincident triggers were consistent with
background. Nevertheless, all triple coincidences were investigated further,
and none was identified as a plausible gravitational wave
inspiral signal. In the rest of this section, we detail the investigations of
the loudest triggers in each science run.

In S4, the loudest coincident trigger in non-playground data was found
in H1 and H2, but not in L1, which \emph{was} in operation at that time. This
candidate has a combined SNR of 22.3 and $P_B=42\%$.
The search produced many triggers in both H1 and H2 at this time, reflecting a
transient in the data produced by sharp changes in ambient magnetic fields due
to electric power supplies. The magnetic fields coupled to the suspended test
masses through the magnets used for controlling their position and alignment.
The transients were identified in voltage monitors, and in magnetometers in
different buildings. The transients were rare, and were identified only in
retrospect, when following up the loudest candidates, so they were not used as
data quality vetoes in this analysis.

In the playground data set, there was a louder candidate which has a combined
SNR of 26.6 and $P_B=77\%$. This candidate was recorded during a time with
elevated dust levels (due to proximate human access to the optics enclosure),
which increases the transient noise in the detectors. 
Therefore this candidate was not considered to be a plausible \GW event.

In S3, the loudest candidate was found in coincidence in H1 and H2, but not in
L1, which \emph{was not} in operation at that time. This candidate has a combined SNR
of 107, resulting from a SNR of 156 in H1 and 37 in H2. It lies
above all background triggers and therefore has less than one
percent probability of being background. None of the
auxiliary channels of the Hanford observatory show suspicious behavior at this
time. This event was a plausible candidate and warranted further investigations
via various follow-ups to confirm or reject a detection. 

\begin{figure}[b]
\vspace{.1cm}
\includegraphics[width=1.65in,height=1.8in]{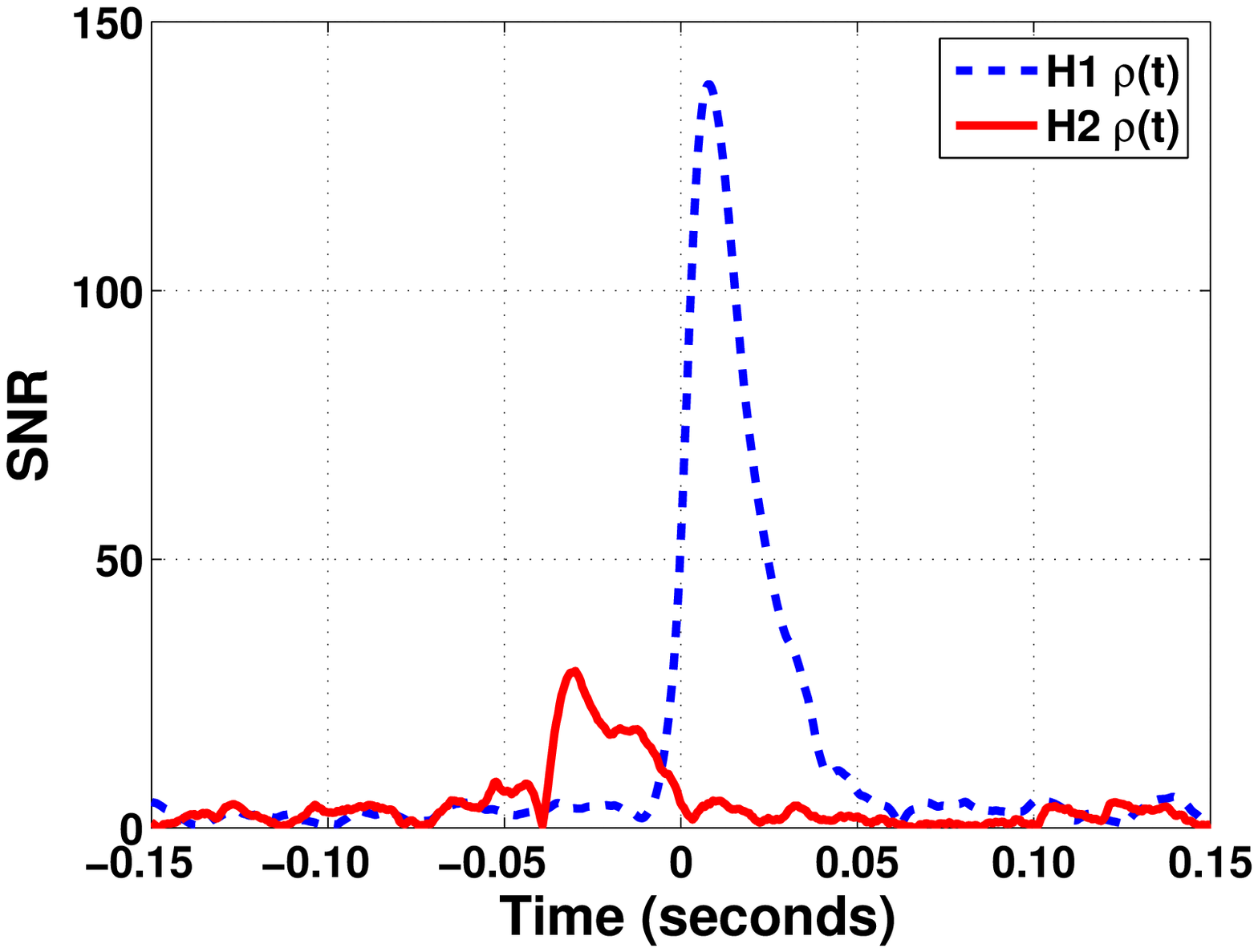}
\includegraphics[width=1.65in,height=1.8in]{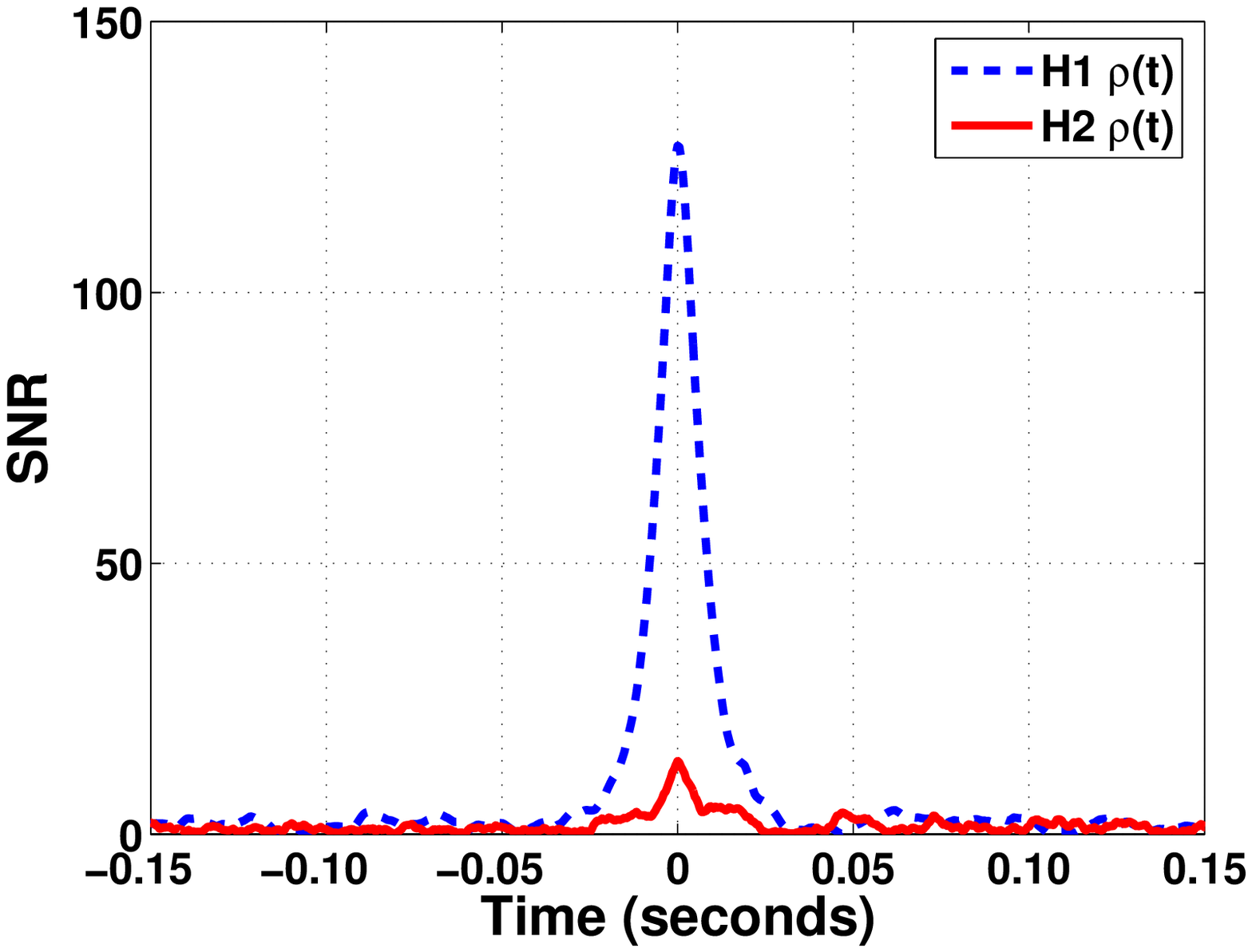}
\caption{Time offset between the H1 and H2 SNR time series, using the same
template. Around the loudest candidate found in the S3 BBH search (left panel),  the maximum of the H2 SNR time series is offset by 38~ms with respect to the maximum of the H1 SNR time series, which is placed at zero time in 
this plot. In contrast, simulated injections of equivalent \GW waveforms
with the same SNR and masses give a time-offset distribution centered around
zero with a standard deviation about 6.5~ms (right panel).} 
\label{fig:S3BBH2}
\end{figure}

We re-analyzed the segment at the time of this candidate with physical
template families. At the coincidence stage, very wide
coincidence windows in time ($\pm25~ms$) and chirp mass ($\pm4~M_\odot$), 
were required to get a
coincident trigger. Then, based on the parameters of this coincident trigger, 
we analyzed the H1 and H2 data around the candidate
time with the \emph{same} template.  We compared the H1 and H2 SNR time series;
a real signal would produce a peak with the same
time of arrival in both instruments to good accuracy. As seen in Fig.~\ref{fig:S3BBH2}, the maxima of both SNR time
series are offset by $38~ms$, which is much larger than expected from
simulations of equivalent \GW waveforms with similar SNR and masses.%
Therefore,
we ruled out this candidate from our list of plausible candidates. 

In summary, examination of the most significant S3 and S4 triggers did not
identify any as likely to be a real gravitational wave.

\begin{table*}[htdp]
\caption{Characteristics of the loudest in-time coincident events found in the
entire S4 data sets. Follow-up analysis
of each of these events, described in section~\ref{sec:loudest}, led us to rule
them out as potential \GW detections. Each loudest event was used in the final
upper limit calculations. The first column shows the
search considered. The second column gives the type of coincidence. The
third column gives the combined SNR $\rho_c$. The
fourth column contains the parameters of the templates that
produced the loudest triggers associated with this event. In
the BNS and PBH binary searches, we provide the mass pairs
 $m_1, m_2$ that satisfy coincidence conditions for chirp mass and symmetric
mass ratio. The two masses can be
significantly different because the coincidence condition on $\eta$ is loose.
In the BBH search, we provide the values of $\psi_0$ and $\psi_3$. The fifth
column is the effective distance in each detector which is provided for the BNS
and PBH search only. The last column is the probability that all
background events have a combined SNR less than $\rho_c$.}
\begin{ruledtabular}
\begin{tabular}{clcccc}
 &Coincidence &  $\rho_{c}$ & $m_1, m_2 (M_\odot)$ & $D_{\rm
eff}$(Mpc) &
$P_{\rm B}(\rho_{c})$\\\hline
PBH & (H1-L1) & 9.8&(0.6,0.6) (H1), (0.9,0.4) (L1) & 
7.4 (H1), 0.07 (L1) & 0.58\\ 
BNS & (H1-L1) & 9.1 & (1.6,0.9) (H1), (1.2,1.2) (L1) & 
15 (H1), 14 (L1) &  0.15 \\
\hline
& &   & $\psi_0 ({\rm Hz}^{5/3})$, $\psi_3 ({\rm Hz}^{2/3})$ &  & \\\hline
BBH & (H1-H2) & 22.3 & (29000, -1800) (H1) &  - &0.42\\
BBH & (H1-H2) (playground time) & 26.6&(153000, -2400) (H1) & - &
 0.77 \\

\end{tabular}
\end{ruledtabular}
\label{tab:Candidates}
\end{table*}

\begin{figure}[htdp]
\includegraphics[width=2.8in,height=2in]{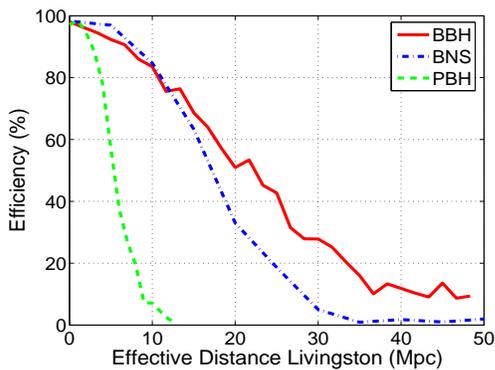}
\caption{Detection efficiency versus effective distance for the
different
searches (S4 run). The BBH and BNS efficiencies are similar, mainly because the
loudest candidate in the BBH search is twice as loud as in the BNS search (See
Table \ref{tab:Candidates}).}
\label{fig:EffvsDist}
\end{figure}
 
\section{Upper limits}\label{sec:ul}
Given the absence of plausible events in any of the six searches
described above, we set upper limits on the rate of
compact binary coalescence in the universe. We use only the results from the more
sensitive S4 data and use only non-playground data in order to avoid
biasing our upper limits through our tuning procedure. The upper limit
calculations are based on the loudest event statistic~\cite{loudestGWDAW03,ul}, 
which uses both the detection efficiency at the combined {\SNR} of the
loudest event and the associated background probability.

The Bayesian upper limit at a confidence level $\alpha$, assuming a uniform
prior on the rate $R$, is given by \cite{ul}
%
\begin{equation}
  1 - \alpha = e^{-R\, T \, \mathcal{C}_{L}(\rho_{c,\mathrm{max}})}  
    \left[ 1 +
 \left( \frac{\Lambda}{1+\Lambda} \right)
    \, R\, T \, \mathcal{C}_{L}(\rho_{c,\mathrm{max}}) \right] 
    \label{bayesianprob}
\end{equation}
%
where $\mathcal{C}_{L}(\rho_{c,\mathrm{max}})$ is the cumulative blue-light luminosity
we are sensitive to at a given value of combined {\SNR} $\rho_{c,
\mathrm{max}}$, $T$ is the
observation time, and $\Lambda$ is a measure of the likelihood that the 
loudest event is due to the foreground, and given by
\begin{equation}\label{xi}
  \Lambda =  
  \frac{|\mathcal{C}_{L}^{\prime}(\rho_{c,max})|}{P_{B}^{\prime}(\rho_{c,max})}
  \left[
  \frac{\mathcal{C}_{L}(\rho_{c,max})}{P_{B}(\rho_{c,max})}
  \right]^{-1}
  \, ,
\end{equation}
where the derivatives are with respect to $\rho_{c}$.  As mentioned in
Sec.~\ref{sec:results}, $P_{\rm B}(\rho)$ is the probability that
all background events have a combined {\SNR}
less than $\rho$ (shown in Table~\ref{tab:Candidates}
for the loudest candidates in each search).  In the case where the loudest event
candidate is most likely due to the background, $\Lambda \rightarrow 0$ and the
upper limit becomes
\begin{equation}\label{ul_bg}
  R_{90\%} = \frac{2.3}{ T \, \mathcal{C}_{L}(\rho_{c,max})} \, .
\end{equation}
In the limit of zero background, i.e. the event is definitely foreground, 
$\Lambda \rightarrow \infty$ and the numerator in Eq.~(\ref{ul_bg}) becomes
$3.9$. The observation time $T$ is taken from Table~\ref{tab:analysedtimes},
where we use the analyzed time {\em not} in the playground.

The cumulative luminosity function $\mathcal{C}_{L}(\rho_{c})$ can be obtained
as follows. We use simulated injections to evaluate the efficiency $\mathcal{E}$
for observing an event with combined SNR greater than $\rho_{c}$, as a
function of the binary inspiral chirp mass $\mathcal{M}_c$ and effective
distance $D_{\rm eff}$. We then integrate $\mathcal{E}$ times the predicted
source luminosity $L(D_{\rm eff},\mathcal{M}_{c})$ as a function of effective
distance and mass. The detection efficiency is different for binary systems of
different masses at the same effective distance. Since we use a broad range of
masses in each search, we should integrate the efficiency as a function of
distance {\em and} chirp mass.  For low mass systems where the
coalescence occurs outside the most sensitive region of the LIGO frequency band,
the distance at which
the efficiency is $50\%$ is expected to grow with chirp mass: $D_{\rm eff,
50\%} \propto \mathcal{M}_c^{5/6}$ (e.g., \cite{thorne.k:1987}).  We can define
a ``chirp distance" for some
fiducial chirp mass $\mathcal{M}_{c,o}$ as $D_c=D_{\rm eff}
(\mathcal{M}_{c,o}/\mathcal{M}_c)^{5/6}$, and then measure the efficiency as a
function of $D_c$ rather than $D_{\rm eff}$.  This efficiency function is now
independent of chirp mass, and the integration can be performed with respect to
the chirp distance only: $\mathcal{C}_{L}~=\int dD_{c} \, \mathcal{E}(D_c) \,
L(D_c)$. We use a model based on \cite{LIGOS3S4Galaxies} for the
distribution
of blue luminosity in distance to calculate $L(D_c)$ for a given mass
distribution (e.g., uniform or Gaussian distribution). Since a system
will have in general slightly different orientations with respect to the two
LIGO observatories, they will also have slightly different effective distances. 
The efficiency for detection is thus a function of both distances, and the
integration needed is two-dimensional:
\begin{widetext}
\begin{equation}\label{eq:cl}
  \mathcal{C}_{L}(\rho)=\int_0^{\infty} \int_0^{\infty} 
  \mathcal{E}(D_{c,H},D_{c,L},\rho)  L(D_{c,H}, dD_{c,L})\, dD_{c,L}
\,dD_{c,H}\,.
\end{equation}
\end{widetext}
The detection efficiency as a function of the effective distance for  each
observatory is shown in Fig.~\ref{fig:EffvsDist}. This efficiency is  computed
using a Gaussian mass distribution, with a mean of
$\mathcal{M}_{c,o}\simeq0.7 M_\odot$ for the PBH binaries ($m_1=m_2=0.75
M_\odot$),  $\mathcal{M}_{c,o}\simeq1.2 M_\odot$ for the BNS
($m_1=m_2=1.4 M_\odot$), $\mathcal{M}_{c,o}\simeq4.4
M_\odot$ for the BBH ($m_1=m_2=5 M_\odot$) and a $1 M_\odot$  standard 
deviation. These efficiencies are
measured with simulated injected signals, using the same pipeline we used to
search for signals; the efficiency is the ratio of the number of injections
detected with SNR above $\rho_{c,{\rm max}}$ to the total number injected. We
show in Fig.~\ref{fig:LumvsDist} the cumulative luminosity as
a function of effective distance in each observatory. It can be seen
that the sharp drop in efficiency in Fig.~\ref{fig:EffvsDist} happens at
approximately the calculated horizon distance shown in
Fig.~\ref{fig:LumvsDist}. 

The upper limit calculation takes into account the possible errors which
arise in a search for {\PBH} binaries and {\BNS}, and are described in some
detail in \cite{systematics}. We follow the analysis presented there to
calculate the errors for the above result. The most significant
effects are due to the possible calibration inaccuracies of the detectors,
(which are estimated by using hardware injections), the finite number of Monte
Carlo injections
performed,  and the mismatch between our search templates and the actual
waveform.  We must also evaluate the systematic errors associated with the
astrophysical model of potential sources within the galaxy described in
\cite{LIGOS3S4Galaxies}. We obtain upper limits on the rate after
marginalization over the
estimated  errors, as described in \cite{systematics}.

In previous result papers (e.g.,
\cite{LIGOS1iul}), we used the Milky Way Equivalent Galaxy (MWEG) unit
which
is approximately $1.7 L_{10}$, where $L_{10}$ is $10^{10}$ times the
blue solar luminosity. In this paper, the merger rate estimates
are normalized to
galactic-scale blue luminosities corrected for absorption with the
underlying
assumption that merger rates
follow the massive star formation rate and the associated blue light
emission.
This assumption is well justified when the galaxies reached by the
detector
are dominated by spiral galaxies with ongoing star formation like the
Milky
Way.

\begin{figure*}[t]
\mbox{
\includegraphics[width=1.8in,height=0.32\textwidth , angle=-90]{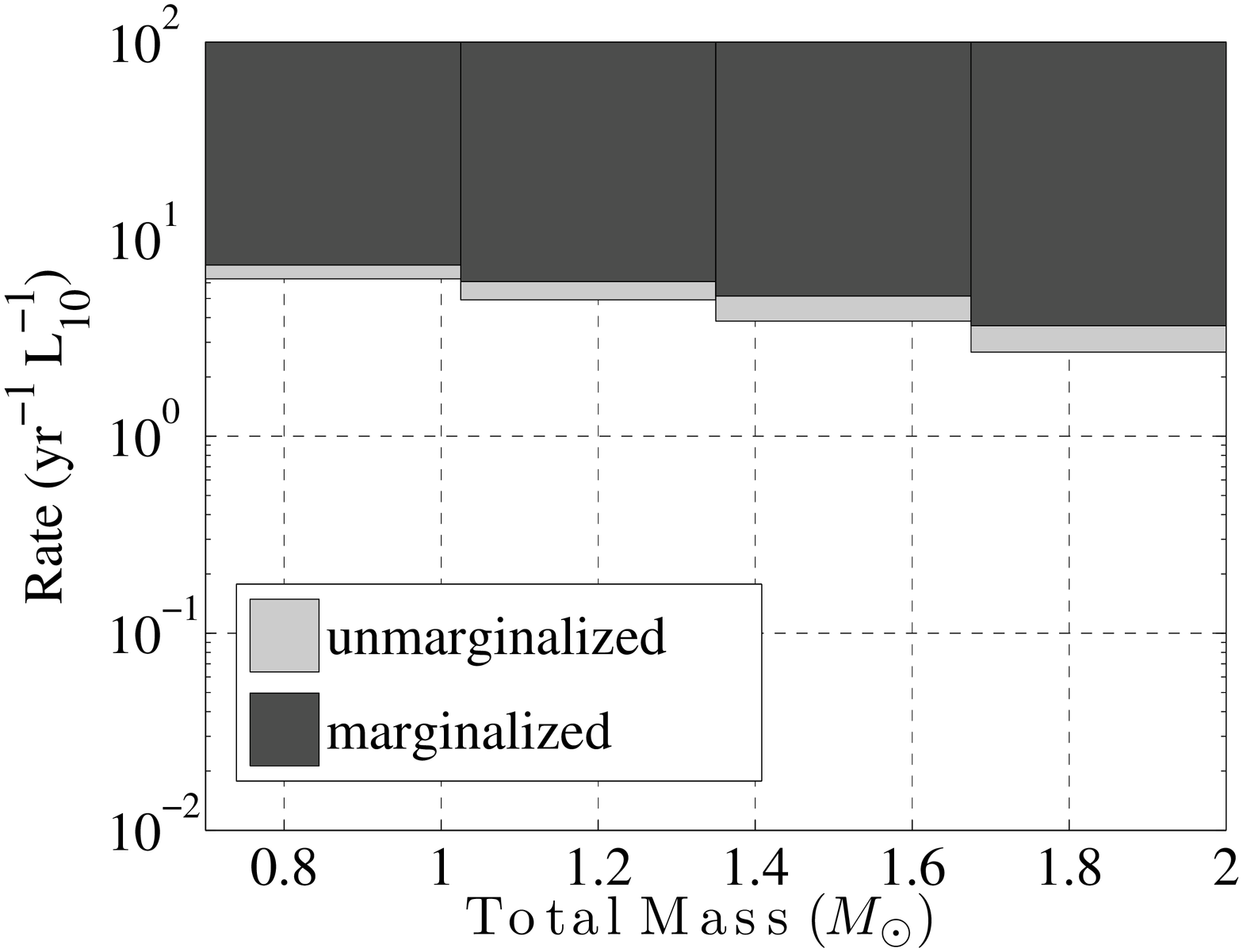}
\includegraphics[width=1.8in,height=0.32\textwidth , angle=-90]{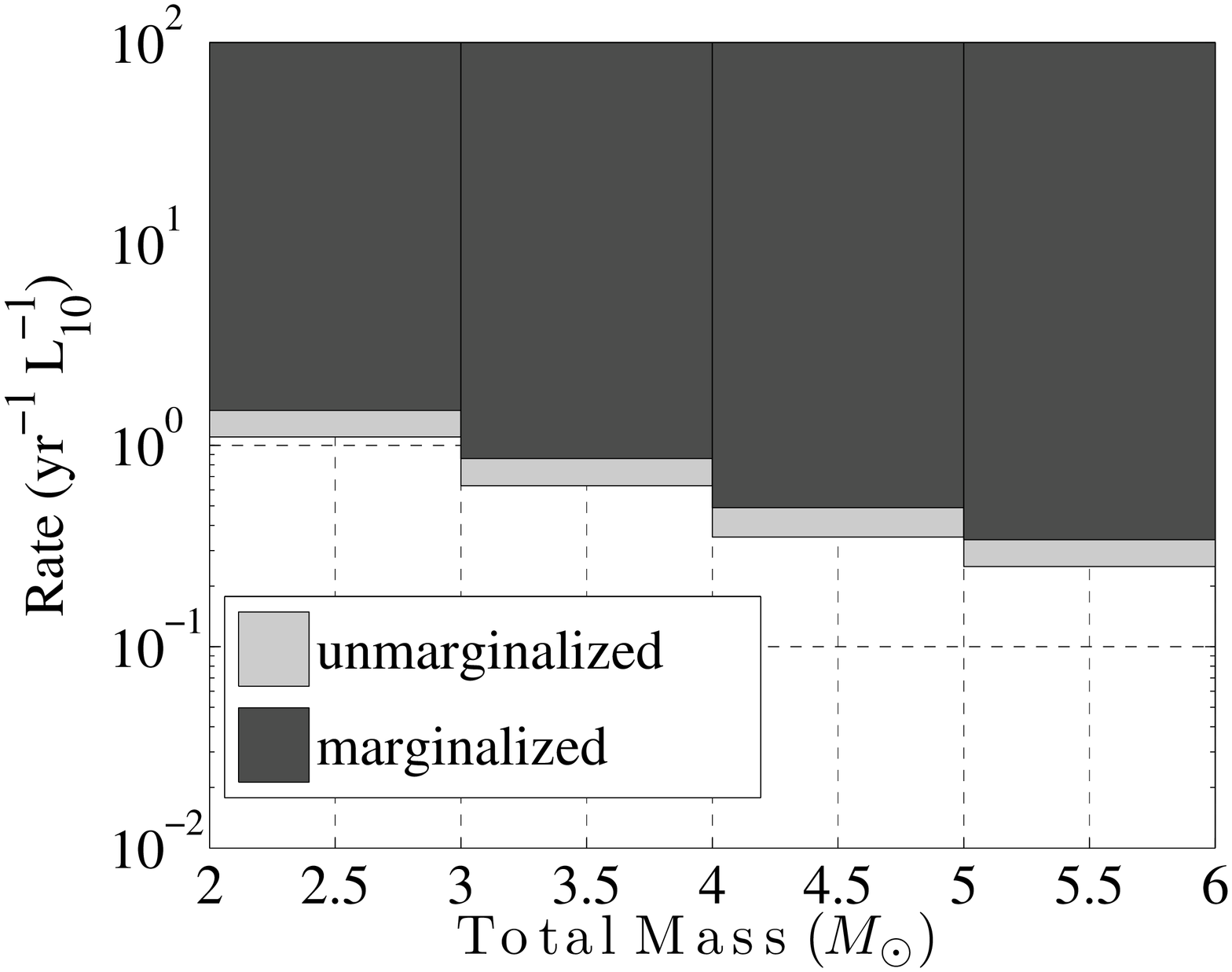}
\includegraphics[width=1.8in,height=0.32\textwidth , angle=-90]{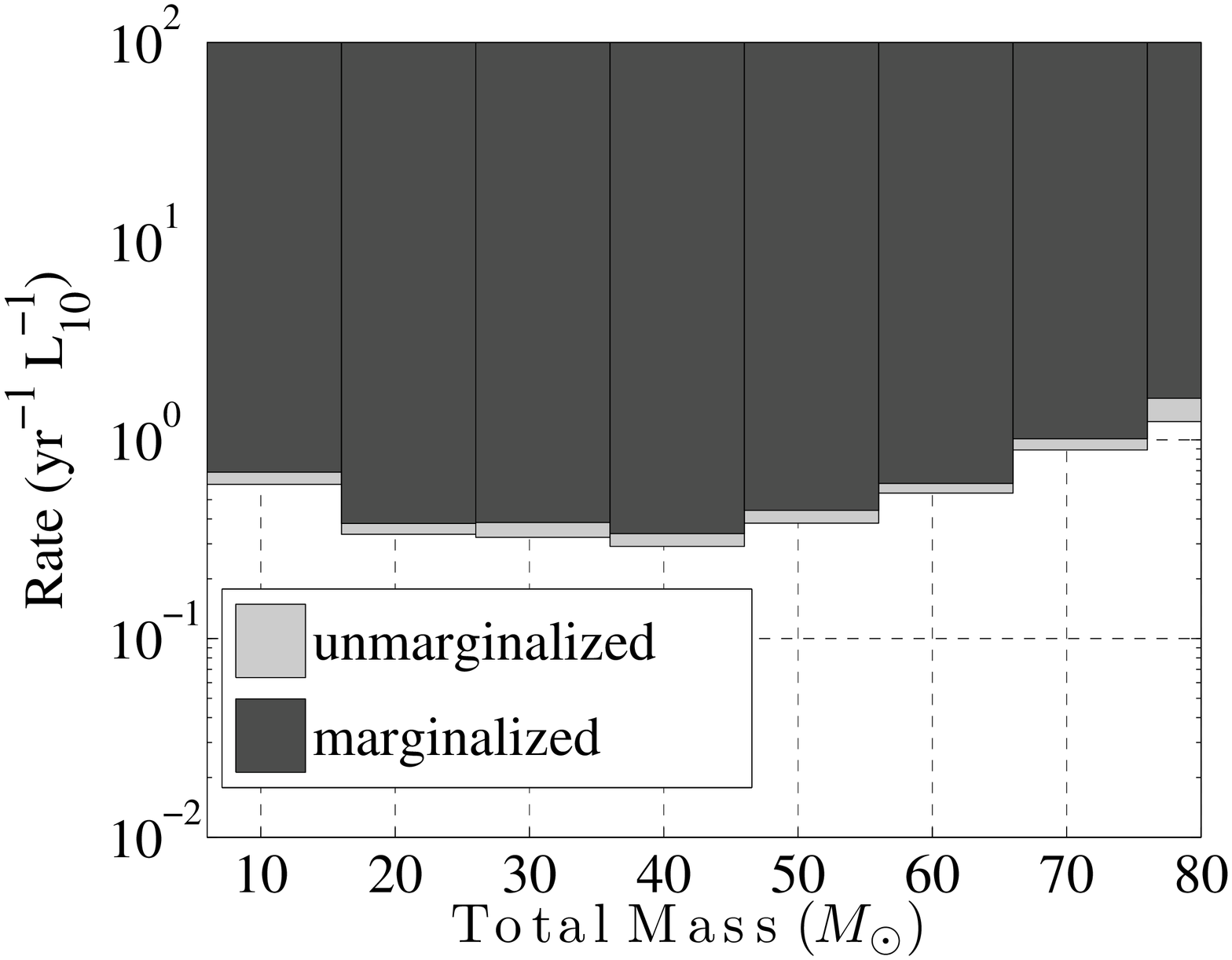}
}
\caption{Upper limits on the binary inspiral coalescence rate per year and per
$L_{10}$ as a function of total mass of the binary, for PBH binaries (left),
BNS (middle), and BBH (right) searches. The darker
area shows the excluded region after accounting for marginalization over
estimated systematic
errors. The lighter area shows the additional excluded region if systematic
errors are ignored. In the PBH binary and BNS searches, upper limits
decrease with increasing total mass, because more-distant sources can be
detected. In the BBH search, upper limits decrease
down to about 30 solar mass and then grow where signals become shorter; this
feature can be seen in the expected horizon distance as well (See
Fig.~\ref{fig:LumvsDist}).
}
\label{fig:upperlimit}
\end{figure*}

Assuming Gaussian mass distributions, as specified above, we obtain upper
limits of $\mathcal{R}_{90\%} = 4.9 \,
\mathrm{yr}^{-1}\,\mathrm{L_{10}}^{-1}$ for {\PBH} binary, $\mathcal{R}_{90\%}
=1.2\,\mathrm{yr}^{-1}\,\mathrm{L_{10}}^{-1}$ for \BNS, and $\mathcal{R}_{90\%}
=0.5 \,\mathrm{yr}^{-1}\,\mathrm{L_{10}}^{-1}$ for \BBH. We also calculated the
upper limits as a function of total mass of the binary, from $0.7~M_\odot$ to
$80~M_\odot$. These upper limits are summarized in Fig.~\ref{fig:upperlimit}.

For comparison, we review the limits on the compact binary coalescence 
rates from previous searches. The best previous limits were obtained 
by TAMA300 using 2705 hours of data taken during the years 2000-2004; 
their result was an upper limit on the rate of binary coalescences in 
our Galaxy of $20\,\mathrm{yr}^{-1}\,\mathrm{MWEG}^{-1}$ ($12\,\mathrm{yr}^{-1}\, \mathrm{L_{10}}^{-1}$)~\cite{TAMA:2006}. Previous
$90\%$ limits from LIGO searches were $47\,\mathrm{yr}^{-1}\,\mathrm{MWEG}^{-1}$ ($28\,\mathrm{yr}^{-1}\,\mathrm{L_{10}}^{-1}$) in the 
mass range $[1-3]M_\odot$ \cite{LIGOS2iul},  and $38\,\mathrm{yr}^{-1}\,\mathrm{MWEG}^{-1}$ ($22\,\mathrm{yr}^{-1}\,\mathrm{L_{10}}^{-1}$) in 
the mass range $[3-20]M_\odot$ \cite{LIGOS2bbh} (the numbers in 
brackets are in units per year per $L_{10}$).

\section{Conclusion}

We searched for gravitational waves emitted by coalescing compact binaries in
the data from the third and fourth LIGO science runs. The search encompassed 
binary systems comprised of primordial black holes,
neutron stars, and black holes. The search techniques applied to these
data represent significant improvements over those applied
to data from the second LIGO science run \cite{LIGOS2iul,LIGOS2macho,LIGOS2bbh}
due to various signal consistency tests which have significantly reduced the
background rates at both single-detector and coincidence levels. Simulated
injections with SNR as low as 8 are detectable, extending the range of
detection. In addition, the stationarity and sensitivity of the data from the S3
and S4 runs were significantly better than in S2. In the 788 hours
of S3 data and 576 hours of S4 data, the search resulted in no plausible
gravitational wave inspiral events. 

In the absence of detection, we calculated upper limits on compact
binary coalescence
 rates. In the {\PBH} binary and {\BNS} searches, the upper limits are close to
values estimated using only the sensitivity of the detectors and the amount of
data searched. Conversely, in the
{\BBH} search, the short duration of the in-band signal waveforms and the
absence of $\chi^2$ veto  resulted in a significantly higher rate of background
events, both at the single-detector level and in coincidence. Consequently, we
obtained a reduced detection efficiency at the combined SNR of the loudest
events and therefore a worse upper limit than we would have obtained
using more effective background suppression, which is under development. The
upper limits, based on our simulations and the loudest event candidates,
are $\mathcal{R}_{90\%} = 4.9,\, 1.2,\, \textrm{and}\,0.5
\,\mathrm{yr}^{-1}\,\mathrm{L_{10}}^{-1}$ for {\PBH} binaries, {\BNS}, and
{\BBH}, respectively. These upper limits are still far away from the theoretical
predictions (see Sec.~\ref{sec:overview}). For instance, the  current
estimate of {\BNS} inspiral rate is $10\textrm{--}170 \times 10^{-6}
\textrm{yr}^{-1} L_{10}^{-1}$.

We are currently applying these analysis methods (somewhat improved) to data
from LIGO's fifth science run (S5). In S5, all
three detectors have achieved their design sensitivity and one year of
coincident data are being collected. We also plan to use physical template
families in the {\BBH} search so as reduce the background and increase our
confidence in detection. In the absence of detection in S5 and future
science runs, the upper limits derived from the techniques used in this
analysis are expected to be several orders of magnitude lower than those
reported here.

\acknowledgments
The authors gratefully acknowledge the support of the United States
National Science Foundation for the construction and operation of the
LIGO Laboratory and the Science and Technology Facilities Council of the
United Kingdom, the Max-Planck-Society, and the State of
Niedersachsen/Germany for support of the construction and operation of
the GEO600 detector. The authors also gratefully acknowledge the support
of the research by these agencies and by the Australian Research Council,
the Council of Scientific and Industrial Research of India, the Istituto
Nazionale di Fisica Nucleare of Italy, the Spanish Ministerio de
Educaci\'on y Ciencia, the Conselleria d'Economia, Hisenda i Innovaci\'o of
the Govern de les Illes Balears, the Scottish Funding Council, the
Scottish Universities Physics Alliance, The National Aeronautics and
Space Administration, the Carnegie Trust, the Leverhulme Trust, the David
and Lucile Packard Foundation, the Research Corporation, and the Alfred
P. Sloan Foundation. This paper has been assigned LIGO Document Number
LIGO-P060045-04-Z.

\bibliography{paper.bbl}

\end{document}